\documentclass{optica-article}

\journal{opticajournal} 

\articletype{Research Article}

\usepackage{lineno}

\usepackage{dcolumn}
\usepackage{bm}
\usepackage[dvipsnames]{xcolor}
\usepackage{amsmath}
\usepackage{mathtools}
\usepackage{braket}

\begin{document}

\title{Integrated photonic platform with high-speed entanglement generation and witnessing}

\author{Gong Zhang\authormark{1,2,\dag}, Chao Wang\authormark{1,\dag}, Koon Tong Goh\authormark{1}, Si Qi Ng\authormark{1}, Raymond Ho\authormark{1}, Henry Semenenko\authormark{3}, Srinivasan Ashwyn Srinivasan\authormark{3,4}, Haibo Wang\authormark{1}, Yue Chen\authormark{1}, Jing Yan Haw\authormark{1}, Xiao Gong\authormark{1}, Joris Van Campenhout\authormark{3}, Charles Lim\authormark{1,5,*}}

\address{\authormark{1}Department of Electrical \& Computer Engineering, National University of Singapore, Singapore\\
\authormark{2}State Key Laboratory of Extreme Photonics and Instrumentation, Zhejiang Key Laboratory of Optoelectronic Information Technology, College of Optical Science and Engineering, Zhejiang University, Hangzhou, China\\
\authormark{3}IMEC, Kapeldreef 75, 3001 Heverlee, Belgium\\
\authormark{4}Lightmatter Canada Inc., 18 King Street East, Suite 1200, Toronto, Canada\\
\authormark{5}Centre for Quantum Technologies, National University of Singapore, Singapore\\
\authormark{\dag}The authors contributed equally to this work.}
\email{\authormark{*}charleslim.research@gmail.com}

\begin{abstract} 
High-speed generation and efficient entanglement detection on a photonic chip are essential for quantum information applications but hard to achieve due to common photonic chips' material properties and limited component performance. In this work, we experimentally demonstrate entanglement witness on a silicon photonic chip, with multi-rail single-photon entanglement generation based on decoy-state techniques. The detection is based on balanced homodyne detectors on the same photonic chip with a bandwidth of up to 12.5 GHz, which allows room-temperature operation. A loss-equivalent analysis method compensates for optical losses and system noises. Experimental results quantify an entangled state fidelity of 92\% in quantum state tomography and a Clauser-Horne-Shimony-Holt (CHSH) violation lower bound of 2.59. These results establish a viable path toward fully integrated, high-bandwidth, room-temperature quantum photonic systems, with potential applications in on-chip quantum optics and quantum random number generation.
\end{abstract}

\section{\label{sec:intro}Introduction}

Integrated photonics technology provides a promising approach for scalable quantum information processing, with appealing features of cost-effectiveness, robustness, and high-speed signal processing~\cite{pelucchi2022potential,slussarenko_photonic_2019,wang_integrated_2019}. The fundamental building blocks to this goal are high fidelity generation, precise manipulation, and efficient measurement of quantum states. Specifically, the entangled quantum state attracts a particular interest~\cite{chen2021quantum}, as quantum entanglement is a primary element for the fundamental study of quantum mechanics, and lies at the core of many quantum information applications which show unique advantages over their classical counterparts. For example, quantum communication~\cite{xu2020secure,luo2023recent}, quantum computing~\cite{abbas2024challenges,couteau2023applications,masada2015continuous}, quantum imaging and sensing~\cite{demille2024quantum,pirandola2018advances,lemos2014quantum}, etc. 

Up to now, integrated photonic platforms based on different materials and techniques have been developed to generate entangled photonics states, including optical nonlinearity, quantum dots, and quantum gates~\cite{wang201818,zeuner2021demand,jin2022generation,zheng2023multichip,steiner2021ultrabright }. However, there is still a lack of experimental realization that could integrate all the critical components on a single photonic chip. Even the state-of-the-art quantum computer uses modular sources, switches, and detectors, with optical fiber interconnections between chips \cite{aghaee2025scaling}. Full integration is important because only system-level integration could unleash the full advantage of integrated quantum photonics in terms of cost-effectiveness and scalability. 

One technological challenge towards this target is that integrated photonics platforms have unique strengths and weaknesses in entanglement generation and detection. Integrating various platforms on the same chip to combine the advantages is generally difficult. Spontaneous parametric down-conversion (SPDC) and spontaneous four-wave mixing (SFWM) are widely deployed for entanglement generation, based on the second-order ($\chi^{(2)}$) or third-order ($\chi^{(3)}$) optical nonlinearity, respectively. Leveraging the mature silicon-on-insulator (SOI) technology, silicon photonics is an appealing approach with the capability of mass manufacturing, high-speed signal processing, and compatibility with Complementary Metal Oxide Semiconductor (CMOS) manufacturing~\cite{yard2024chip,tasker2024bi,chen2024ultralow}. However, limited by strong two-photon absorption for near-infrared (NIR) wavelengths, silicon typically has a relatively limited entanglement generation rate via SFWM~\cite{liu2018influences,silverstone2015qubit,husko2013multi}. Lithium niobate features a strong nonlinearity for efficient photon-pair generation based on SPDC~\cite{javid2021ultrabroadband,xue2021ultrabright,mondain2019chip, yadav2024high, zhuang2025ultrabright}, but the material itself does not offer photon detection mechanisms for quantum state measurement~\cite{zhu_integrated_2021}. Enormous efforts have been made to integrate different material technologies into a single photonic chip or package, towards the system's full functionality with an uncompromising performance for each photonic element. To this end, additional technological investigations are necessary to overcome the fabrication challenge, for example, the lattice mismatch between materials, and to develop simple fabrication processes suitable for mass production~\cite{komljenovic2018photonic,kaur2021hybrid}.

Another hurdle to a fully integrated quantum system is the limited availability or performance of some critical components with current technologies. One example is the integration of single-photon detectors with photonic integrated circuits. Intensive endeavors have been made to achieve this goal, with recent demonstrations on integrated circuits with single-photon avalanche diodes (SPAD)~\cite{zhang2023hybrid,vines2019high,yanikgonul2019simulation} and superconducting nanowire single-photon detectors (SNSPD)~\cite{gyger2021reconfigurable,wolff2021broadband,akhlaghi2015waveguide}. However, they have not been widely adopted yet for full system integration, as further investigations are needed to improve the detector performance, and tackle the photon processing issue at cryogenic temperatures~\cite{Lange:22,eltes_integrated_2020}. A promising alternative is the balanced homodyne detector (BHD) for continuous variable systems~\cite{ng2024chip,thearle2018violation, bruynsteen2021integrated,zhang2019integrated}, which also shows great potential for detecting single photons \cite{sidhu2025security,qi2020characterizing}. It requires only photodiodes and amplifiers working at room temperature. However, homodyne detectors are sensitive to phase noise and vaccum states. Thus, protocol and system-level design are required for the application in entanglement witnessing.  Another example is the stringent requirement for optical filtering in integrated quantum systems. To obtain high-quality entangled photon generation, it often requires narrow-band filtering on the generated photo pairs or delicate engineering of the optical nonlinear processes~\cite{lu_advances_2021,Meyer-Scott:18}. Moreover, efficient pump filtering is also needed to isolate the signal and idler photons from strong pump noises, which is especially challenging for SFWM where all the involved fields have a similar wavelength~\cite{caspani_integrated_2017,arrazola_quantum_2021}. 

In this paper, we report an experimental demonstration of quantum entanglement witnessing with the states generated and measured on a silicon photonic chip using off-chip laser source. 
Specifically, the quantum entangled states are generated by impinging single-photon states on a 50:50 optical beam splitter, leading to the quantum correlations between the Hilbert spaces of the two optical output modes~\cite{fabre_modes_2020}. 
By adopting the decoy-state method from quantum cryptography~\cite{hwang_quantum_2003,lo_decoy_2005,wang_beating_2005,yuan_simulating_2016}, we show that it is sufficient to generate the multi-rail single-photon entangled states using phase-randomized coherent states with several intensity levels. 
Furthermore, we have developed high-speed BHDs on a silicon photonic chip and introduced an innovative loss-equivalent analysis method aimed at removing the stringent requirements for high photodiode efficiency and low electronic noise in the amplifier. The use of single-particle encoding and homodyne detector decoding allows room-temperature and high-bandwidth operation. 
In addition, we provide theoretical proof of the fair sampling condition in our system, ensuring an accurate witnessing of the on-chip quantum entanglement based on the CHSH violation game with post-selected data~\cite{wenger2003maximal, oudot2024realistic}. 
Our experimental results show that our scheme could provide CHSH violation approaching its ideal value ($2\sqrt{2}$) for a maximally entangled state with practical light sources and noisy high-speed homodyne detectors on the same silicon photonic chip. 

The paper is organized as follows. 
In Section.~\ref{sec:entanglementgenerationandmeasurement}, the entanglement generation and measurement scheme is defined and introduced. In Section.~\ref{sec:chipdesign}, the photonic chip design and characterization results are presented. In Section.~\ref{sec:tomo} and Section.~\ref{sec:chsh}, a quantum state tomography based on our setup is demonstrated, and the Bell violation based on CHSH inequality is shown to verify the generated states. 

\section{\label{sec:entanglementgenerationandmeasurement}Entanglement generation and measurement}

First, the working principles of our scheme are explained in detail, including the probabilistic generation of the entangled state and the measurement technique used to achieve near-ideal CHSH violation with imperfect photodiode efficiency and electronic amplifier noise. 

\begin{figure}[tbp]
\centering
\includegraphics[width=0.5\columnwidth]{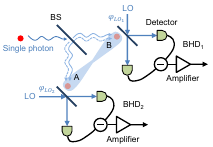}
\caption{(a) Illustration of single-photon entanglement. A single photon impinges onto a 50:50 beam splitter (BS), generating a Bell state in the output two-mode system sharing a single photon. Quantum state measurement consists of two locally operated balanced homodyne detectors (BHD), with a controllable local oscillator (LO) phase $\varphi_{LO_1}$ and $\varphi_{LO_2}$, respectively.  }
\label{fig_entanglement}
\end{figure}

When a single-photon state $\ket{1}$ impinges onto a beam splitter (BS) with a reflection and transmission ratio of 50:50, it generates a Bell state in the output two-mode system
\begin{equation}
\label{eq:singlephotonentanglement}
\ket{\Psi} = \frac{1}{\sqrt{2}} \bigg(\ket{0}_A\ket{1}_B+\ket{1}_A\ket{0}_B\bigg),
\end{equation}
where $A$ and $B$ refer to two output spatial modes sharing the delocalized photon, as illustrated in Fig.~\ref{fig_entanglement}~(a).

To generate single-photon states using practical light sources, the decoy state technique is adapted to simulate the input of pure single photon state~\cite{lo_decoy_2005,wang_beating_2005,yuan_simulating_2016}. 
Specifically, the phase-randomised coherent states $\rho_{\mu}$ comprises a mixture of Fock states
\begin{equation}
\label{eq:phaserandomisedWCS_new}
\rho_{\mu} = \int_{0}^{2\pi} \frac{1}{2\pi}\ket{\sqrt{\mu}e^{i\theta}}\! \bra{\sqrt{\mu}e^{i\theta}} \,d\theta =  \sum_{n} \frac{\mu^n}{n!} e^{-\mu}\ket{n}\!\! \bra{n},
\end{equation}
where $\theta$ is the phase of coherent state, and $\ket{n}$ represents a Fock state that contains $n$ photons for $n\in\mathbb{N}$. $\mu^n e^{-\mu}/n!$ is the photon-number distribution of coherent states, following a Poisson distribution. Here, in our experiment, we would like to extract the single-photon component.

The input is a mixture of phase randomised coherent states $\rho_{\mu}$ with $L$ different intensities $\mu \in \{\mu_{0,1,2,...,L} | \mu_0<\mu_1<\cdots<\mu_L\}$. 
The phases of prepared states are assumed to be random and cannot be accessed by the measurement devices. The random phase eliminates the coherence among Fock states and enables the analysis of single-photon entanglement, in conjunction with the decoy-state method. Therefore, single-photon entanglement can be extracted from an ensemble devoid of quantum correlations. We believe the assumption is reasonable in our scenario since we are not running a quantum key distribution (QKD) protocol, and the user has full control of the chip.

Considering a general measurement device, the measured probability (or probability distribution) of the input state $\rho_{\mu}$ is defined as the gain 
\begin{equation}
\label{eq:Qmu}
Q_\mu = \sum_{n} P(n) \frac{\mu^n}{n!} e^{-\mu},
\end{equation} 
where $P(n)$ is the measured probability (or probability distribution) from input Fock states $\ket{n}$. Following previous paper~\cite{lo_decoy_2005,yuan_simulating_2016}, we define $P(n)$ as the yield $Y_n$. Since the gain $Q_\mu$ is a linear combination of yield $Y_n$, the contribution of single-photon state $Y_1$ can be precisely estimated by solving a set of linear equations, using experimental statistics as constraints and coherent states with various intensity levels.

To measure such a two-mode entangled state is a non-trivial task. Here, we use two locally operated BHDs for quantum state characterization and the CHSH Bell test for entanglement witnessing. 
The BHD measures the quadrature of the corresponding input optical mode. When used with local oscillators (LOs) possessing a well-defined phase reference, the BHD pair provides an efficient method for characterizing the delocalized entangled state~\cite{babichev_homodyne_2004,morin_witnessing_2013,fuwa_experimental_2015}. We provide details about the quantum state tomography in Sect.~\ref{sec:tomo} for the single-photon entangled state characterization generated in our system.

For entanglement witnessing, we further examined it using the CHSH Bell test. To apply the Bell theorem to the continuous field quadrature, we follow the approach outlined in Ref.~\cite{babichev_homodyne_2004} and provide proof that the fair-sampling condition holds in our system. 
The homodyne detector outputs $x$ are post-selected with a fixed binning threshold $T$, such that the $0,1$ results are assigned for $x<-T$ and $x>T$, respectively. As such, the CHSH violation faithfully witnesses the quantum entanglement generated on the chip under the condition where the quantum states are restricted within the $\{\ket{0},\ket{1}\}$ subspace. The details of the proof can be found in Supplementary Information~S1.

Practical imperfections, including propagation loss, detector inefficiency, and electronic noise, could spoil the quantum state in the experiment and ruin the entanglement witnessing. To mitigate these effects in the CHSH Bell test, we come up with a loss-equivalent analysis method, which combines the modeling of the realistic BHDs based on assumptions on its noise property and a modification in the decoy-state method.

\begin{figure*}[tbp]
\centering
\includegraphics[width=0.9\columnwidth]{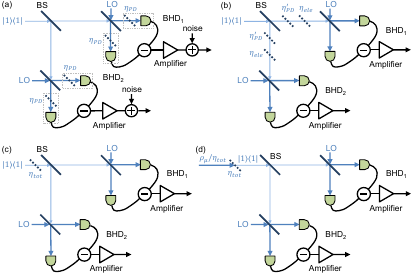}
\caption{Schematic of the loss-equivalent scheme (a) Actual experimental setup. Photodiodes in BHD have non-unit efficiencies, and the electronic noise in BHD is assumed to be independent of the measured optical signal and possesses a Gaussian dissolution. (b) Both the inefficiency of photodiodes~\cite{leonhardt_realistic_1993} and electronic noise~\cite{appel_electronic_2007} in BHD could be modeled as an optical loss on the input signal. (c) Equivalent scheme with a total optical loss in front of BS. (d) Simulating single photon generation after the equivalent total optical loss with modified decoy-state intensities.}
\label{fig_equivalentscheme}
\end{figure*}

The efficiency of photodetectors is modeled as optical losses (BS with a transmittance of $\eta_{PD}$) in front of the photodetectors, as shown in Fig.~\ref{fig_equivalentscheme} (a). Such a loss model is further equivalent to adding an attenuator (BS with a transmittance of $\eta_{PD}'$) to the signal before mixing it with LO~\cite{leonhardt_realistic_1993}, as shown in Fig.~\ref{fig_equivalentscheme} (b). Besides, suppose it is reasonable to assume that the electronic noise is independent of the measured optical signal with a Gaussian distribution. In that case, its effect on homodyne measurement can also be seen as equivalent to an optical loss (BS with a transmittance of $\eta_{ele}$)~\cite{appel_electronic_2007}. 

Similarly, two BHDs are connected to the two output modes of the BS, the fictitious optical loss or partial beam splitters could be combined, and the scheme is equivalent to placing all optical loss (BS with a transmittance of $\eta_{tot} = \eta_{PD}'\cdot\eta_{ele}$) in front of the BS, as shown in Fig.~\ref{fig_equivalentscheme} (c). Here, we assume the efficiency and electronic noise of two BHDs are the same. If they are unbalanced, the splitting ratio of the BS can be fine-tuned to match the difference.

To further eliminate the effect of optical loss on the input states, the intensity of input phase randomized coherent states is increased by $1/\eta_{tot}$. Note that the loss applied to coherent states only attenuates its intensity. As such, the single-photon states generated using the decoy-state technique after the fictitious optical loss will not be affected, which shall not be confused with the loss analysis in Ref.~\cite{leonhardt_realistic_1993}. As a result, if the detector efficiency and electronic noise calibrations can be trusted, the single photon generation can be simulated without being affected by all the system imperfections. The detection can be considered as using a detector with unit detector efficiency and no electronic noise, as illustrated in Fig.~\ref{fig_equivalentscheme} (d). 

There are three advantages of the proposed scheme. 
First, the loss-equivalent scheme removes the stringent requirements on high efficiency and low noise measurement in the quantum system, allowing almost perfect entanglement witnessing with realistic lossy and noisy components. Secondly, as a result, the critical components can be integrated into a single silicon photonic chip. Beam splitters with tunable splitting ratios can be designed with directional couplers or multimode interferometer (MMI) couplers. High-speed optical modulators and photodetectors are also available in the process design kits (PDKs), for photon modulation and detection, respectively. Third, by accepting a larger electronic noise, one can ease the bandwidth restriction in homodyne detector design, which enables a much faster quantum state measurement~\cite{laudenbach_continuous-variable_2018,lvovsky_continuous-variable_2009}. 

We note that, unlike works aiming for a loophole-free Bell test with minimal assumptions, this work relies on specific assumptions about the experimental setups to simplify the requirements for generating and witnessing quantum entanglement on integrated photonic chips. 

There are several key assumptions in this scheme. Firstly, we assume that the source and detectors conform to the models presented in this paper. Specifically, we assume the source is a phase-randomized coherent state with precise intensity control. On the receiver side, the homodyne detectors are assumed to perform quadrature measurements with precise characterization of quantum efficiency and electronic noise levels throughout the entire duration of the experiment. 
Additionally, we assume the identical and independent (i.i.d.) operation of the system, and our data analysis is based on the asymptotic resource assumption. 

\section{\label{sec:chipdesign}Photonic chip design and performance calibration}

\begin{figure*}[tbp]
\centering
\includegraphics[width=1\columnwidth]{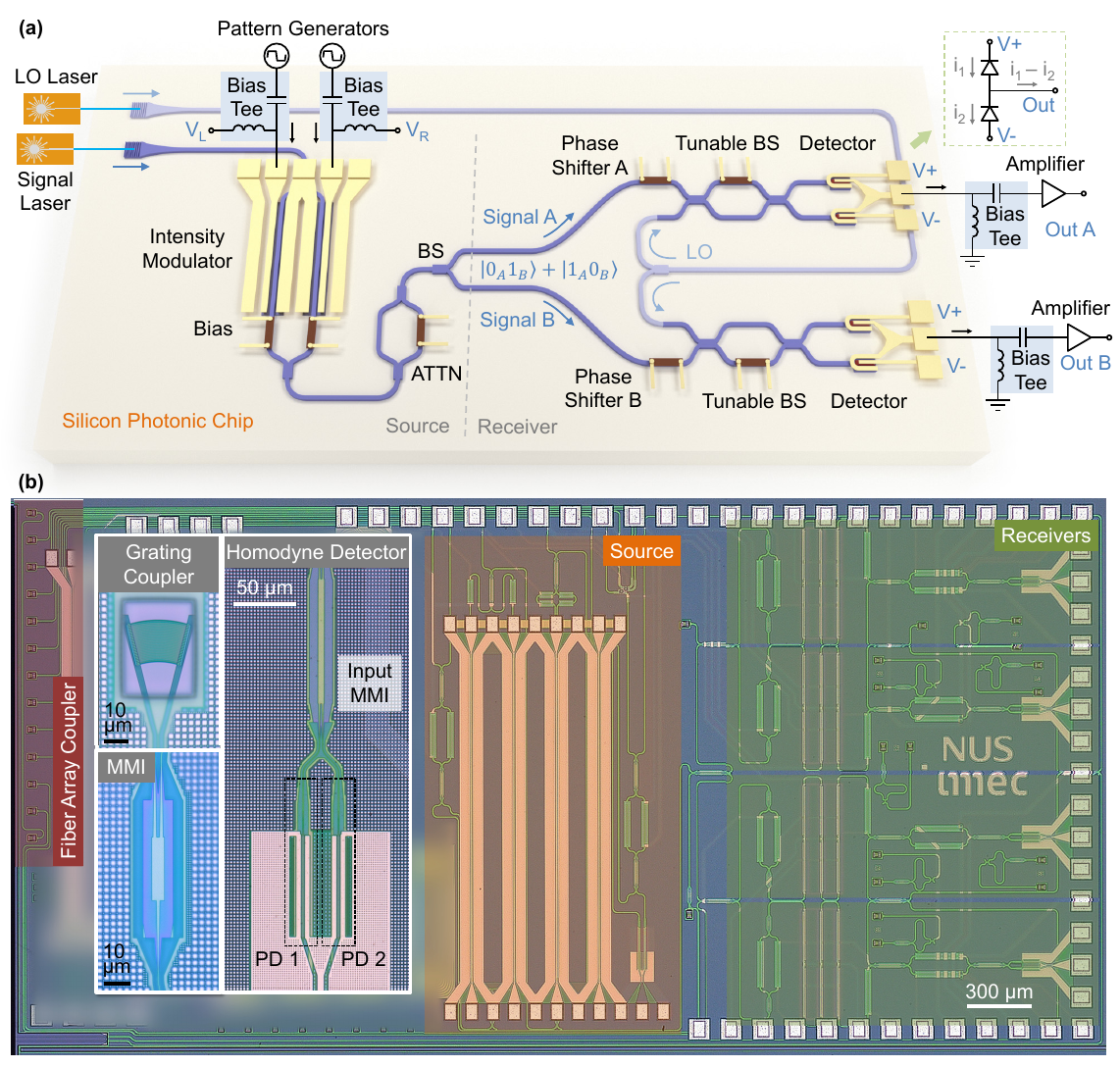}
\caption{(a) The experimental setup for on-chip entanglement generation and witnessing. Two lasers operated near 1550~nm wavelength with about 150~MHz detuning are used as signal and local oscillator (LO). The source part modulates the input signal using an intensity modulator and attenuates the signal to the required single photon energy level. A pattern generator with bias tees is used to drive the modulator. A beam splitter (BS) splits the input states and generates the single photon path-entangled states 
$\ket{\Psi} = (\ket{0}_A\ket{1}_B+\ket{1}_A\ket{0}_B)/\sqrt{2}$. The entangled states are then sent to the homodyne detectors for detection. The phase shifters A\&B select the measurement basis, and the tunable BS precisely balances the detectors. The output of the homodyne detector is amplified using a broadband RF amplifier and monitored using an oscilloscope. (b) The microscopic photo of the photonic chip, with key components false-colored, includes the fiber array coupler, edge coupler, source, and receivers. The inset shows the zoom-in photo of a grating coupler, an MMI, and a homodyne detector.}
\label{fig_overview}
\end{figure*}

Our experimental system is fabricated on a silicon photonic chip using a standard foundry process (IMEC iSiPP200) and PDK optical components. The photonic chip design and its microscopic photo are shown in Fig.~\ref{fig_overview}, where the key components are false-colored.

A 1550 nm wavelength continuous wave (CW) laser is used as the signal source and coupled to the photonic chip via an input grating coupler from the left side of the chip. A carrier depletion Mach-Zehnder modulator (MZM) 
modulates the intensity of the input laser into four levels to apply the decoy state technique. Each driving signal is slightly biased, using bias tees to match MZM's optimal driving voltage range and achieve optical intensity modulation with the highest possible extinction ratio. A thermo-optical phase shifter inside the MZM selects the modulator working point. Subsequently, the modulated laser passes through a series of variable optical attenuators made of Mach-Zehnder interferometer (MZI) and reaches the single-photon energy level. 

The entanglement is generated by sending the single-photon state through an MMI, acting as a 50:50 beam splitter. Considering single-photon state input, the two output optical modes share an entangled state as specified in Eq.~\ref{eq:singlephotonentanglement}. 
The quantum state is defined based on the final analog-to-digital converter (ADC) sample. The input signal is sampled into small uniform time bins with identical properties. As a result, multiple samples can be collected from each modulation level. Thus, the sample rate relates to the bandwidth of BHD only. The maximum sample rate is twice the BHD bandwidth to avoid aliasing.
Each optical mode is measured by a balanced homodyne detector with a local oscillator (LO). The LO is generated from another laser with 170 MHz detuning from the signal laser to avoid any signal distortion caused by noises below the system's low cut-off frequency of 20 MHz. The detuning is intentional to uniformly and randomly distribute the phase of input states. More detail for phase randomization is provided in Supplementary Information S2.

Each homodyne detector consists of a thermo-optical phase shifter for LO phase a MZI as tunable BS for detector balance, and high-speed photodiodes. Noted here, due to the phase-dependent loss of the high-speed modulator, we do not adopt high-speed basis choosing, which does not affect the novelty of this paper and will be covered in future works. Germanium photodiodes with a high bandwidth of up to 50 GHz and a responsivity of 0.8 A/W are used. The homodyne detector's top and bottom electrodes provide photodiode reverse bias voltage, and the center electrode is for signal output. Depending on the amplifier design, the center electrode might require a bias tee with 0~V DC input for a proper working condition of the photodiodes. Radio frequency (RF) low-noise amplifiers operating up to 10 GHz are used in our testing. The output data is collected on an oscilloscope and post-processed offline, including spectrum whitening and noise filtering. 
In our current experiment, high-speed modulation signals and detector outputs are connected using RF probes, while the remaining electrodes are connected by wire bonding. 
For testing purposes, we designed multiple copies of MZMs and homodyne detectors on the chip. Four homodyne detectors are arranged in parallel such that it can support up to four-dimensional entanglement generation and detection. In the current experiment, we use only two homodyne detectors to measure the two-mode entangled state.

\begin{figure}[tbp]
\centering
\includegraphics[width=0.8\columnwidth]{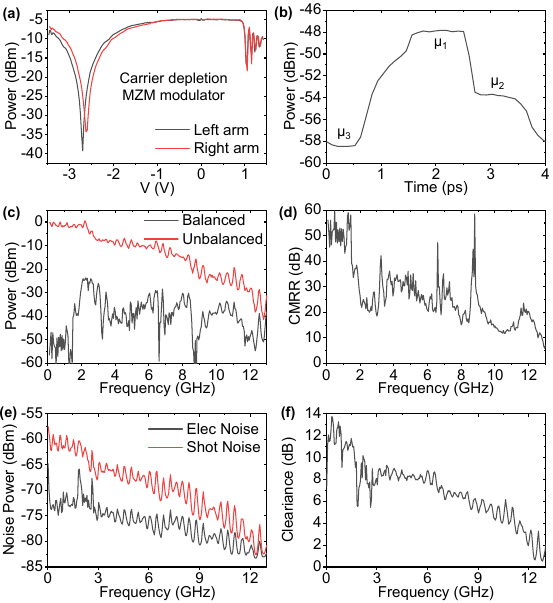}
\caption{Calibration result for chip-based modulators and homodyne detectors. (a) The output power of the carrier depletion MZM modulator as a function of the applied voltage on both arms. (b) The modulated signal is shown in the time domain. The three flat regions are used as three decoy intensities $\mu_1,\mu_2,\mu_3$. (c) Spectrum response for the chip-based homodyne detector when the detectors are balanced and unbalanced. (d) The CMRR result is calculated from (c). (e) The electronic noise spectrum and measured shot noise with about 18~dBm input LO power. (f) The shot noise clearance is calculated from (e).}
\label{fig_mod}
\end{figure}

To have a system calibration of our chip, core components, including chip-based modulators and homodyne detectors, are carefully characterized. Fig.~\ref{fig_mod} (a) shows the optical power as a function of voltage added on both the MZM arms. The result indicates that the maximum extinction ratio can reach close to 35~dB under reverse bias, with $V_\pi$ of about -2.6~V. However, due to the RF crosstalk in our chip (see Supplementary Information~S3), a forward bias driving is chosen to achieve a higher driving efficiency using minimum driving signal power. The resulting modulation speed is limited due to the speed of slow carrier injection and recombination processes in silicon~\cite{liu2008carrier}. Therefore, the modulation speed in our experiment is chosen to be 1~GS/s. As a reference here, the small signal bandwidth of the modulator is 36.8 GHz. The limited modulation speed in our experiment indicates the different requirements of chip components in quantum and classical applications. The optical power as a function of time is measured by sweeping a 50~ps pulsed laser for better timing resolution. Only three of the flat modulation levels are used for the experiment, marked as $\mu_1,\mu_2,\mu_3$, as shown in Fig.~\ref{fig_mod} (b).

For the balanced homodyne detectors, they are first characterized by the common-mode rejection ratio (CMRR), as shown in Fig.~\ref{fig_mod} (c, d). Fig.~\ref{fig_mod} (c) shows the frequency response of the homodyne detector with a modulated LO input. The difference in the output RF power when the two arms are balanced and unbalanced indicates the current subtraction in BHD. Based on our calibration, a 50~dB CMRR is observed below 1.5~GHz, while a 20~dB CMRR can reach up to 6.5~GHz, as shown in Fig.~\ref{fig_mod} (d). 

A faithful implementation of our scheme also requires precise calibration of the detector noise and its shot-noise-limiting performance, which is done by measuring the noise power of the BHD, with and without LO input. 
Noise power calibration results are shown in Fig.~\ref{fig_mod} (e), with a maximum photocurrent of 2.2~mA. Despite some RF noise near the 2.4~GHz band, our homodyne detector achieves more than 10~dB shot-noise clearance below 1.7~GHz, and more than 4~dB up to 10~GHz, as shown in Fig.~\ref{fig_mod} (f). To show the feasibility of the proposed loss-equivalent analysis method, we intentionally reduce the LO photocurrent to 0.4~mA, which increases the portion of electronic noise in the overall output signal, which infers a larger optical loss modeled by the fictitious BS~\cite{appel_electronic_2007}. With the loss-equivalent analysis method implemented, the BHD is considered ideal with unit efficiency and no electronic noise. The output from BHD with only LO input is considered pure shot-noise and is used to define the shot-noise unit throughout the experiment in this manuscript. 
The output data is sampled using an oscilloscope with 12.5~GHz bandwidth, corresponding to a maximum sample rate of 25 GS/s. In our experiment, by keeping only the flat region in the modulated signal, the final sample rate is 8.75~GS/s.

Characterisation of other optical components on the chip can be found in Supplementary Information~S4.

\section{\label{sec:tomo}Quantum state tomography} 
To assess entanglement generated on the chip, quantum state tomography is performed with the decoy-state technique. 
Specifically, we follow the quantum state tomography with the maximum-likelihood approach to construct the density matrix of the quantum entangled state~\cite{lvovsky_continuous-variable_2009}. Different measurement basis for two BHDs are selected by modulating the phase of LOs. 

When a phase-randomized coherent state with intensity $\mu$ is sent through the BS, based on Eq.~\ref{eq:Qmu}, the coincidence detection probability of the two BHDs is defined as
\begin{equation}
\label{eq_q}
Q_{\mu,d\theta}(x_A,x_B) = \sum_{n=0}^{\infty} Y_n (d\theta) \frac{\mu^n}{n!}e^{-\mu},
\end{equation}
where $Y_n (d\theta) \equiv P(x_A,x_B|n,d\theta)$ is the probability density function of outcome $x_A$ and $x_B$ given the input Fock states $|n\rangle$ and measurement phase $d\theta$. The subscript $A, B$ denotes the two output modes $A$ and $B$, respectively.

To get a precise estimation of the yield with single-photon input $Y_1 (d\theta) \equiv P(x_A,x_B|1,d\theta)$, we follow the method by Ref.~\cite{yuan_simulating_2016,valente2017probing}. Considering the input coherent states with $L$ different intensities $\mu_1<\mu_2<\cdots<\mu_L$ and a vacuum states $\mu_0=0$, the yield $Y_{1,d\theta}$ can be estimated by:
\begin{eqnarray}
\label{eq_Y}
Y_{1,d\theta}^{E} =\mu_1 \mu_2 \cdots \mu_L \sum_{j=1}^{L} \frac{\mu_j^{-2}(e^{\mu_j}Q_{\mu_j,d\theta} - Q_{\mu_0,d\theta})}{\prod_{1\leq i\leq L,i\neq j} (\mu_i - \mu_j)},
\end{eqnarray}
where the subscript $E$ denotes an estimation.

\begin{figure}[tbp]
\centering
\includegraphics[width=0.8\columnwidth]{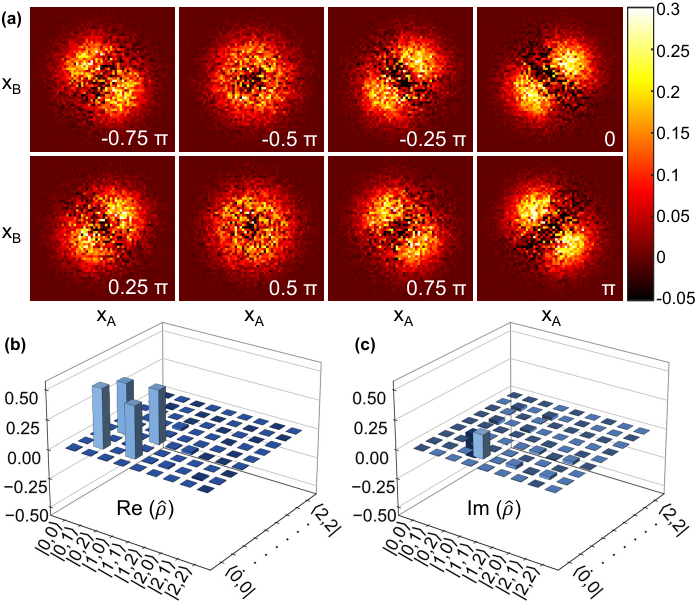}
\caption{(a) The quadrature distribution at different phase difference $d\theta$ measured using decoy state method. The x-axis is the result from Alice, while the y-axis is the result from Bob. The density matrix's real (b) and imaginary (c) parts are calculated using a maximum likelihood solver.}
\label{fig_tomo}
\end{figure}

In this experiment, the three decoy state intensities after compensating the equivalent losses $\mu$ are set to 0.9840, 0.2314, and 0.0872. $10^6$ samples are collected for each decoy state, while $5\times10^7$ vacuum states are collected. The phase difference $d\theta$ between Alice and Bob is scanned from -$\pi$ to $\pi$ with a $\pi/4$ interval. Both the measured quadrature distributions are shown in Fig.~\ref{fig_tomo} (a), showing an apparent correlation behavior. The distributions are then fed to a maximum likelihood solver to obtain the constructed density matrix, as shown in Fig.~\ref{fig_tomo} (b, c). The maximum photon number is cut off at ten photons, and the figure shows only up to two photons for clarity. The complete density matrix is shown in Supplementary Information S5. Based on the density matrix obtained, the fidelity is calculated as 0.9268. The result confirmed that the single-photon entangled state is generated with high fidelity.

\section{\label{sec:chsh}CHSH violation}
We perform the CHSH violation test to verify the generated entanglement further. To this end, two separated BHDs measure optical modes $A, B$ based on randomly modulated phases $\varphi_a\in \{0, \pi/2\}$ and $\varphi_b\in \{-\pi/4, \pi/4\}$, respectively. The outcomes $x_A$ and $x_B$ from their homodyne detectors are post-processed by applying a threshold $T$. The four possible coincidence probabilities can be calculated as
\begin{equation}
\label{eq_p}
\begin{cases}
    P_{00}(a,b) = P(x_A<-T,x_B<-T|a,b) \\
    P_{01}(a,b) = P(x_A<-T,x_B>T|a,b) \\
    P_{10}(a,b) = P(x_A>T,x_B<-T|a,b) \\
    P_{11}(a,b) = P(x_A>T,x_B>T|a,b) \\    
\end{cases} ,
\end{equation}
where the subscript $0,1$ denotes the assigned results when the output is below or above the threshold. The quantum correlation $E$ is then given as 
\begin{equation}
\label{eq_E}
    E(a,b)=\frac{P_{00}+P_{11}-P_{01}-P_{10}}{P_{00}+P_{11}+P_{01}+P_{10}}.
\end{equation}
Then, one can obtain the CHSH violation value $S$ as 
\begin{equation}
    S=E(0,0)+E(1,0)+E(0,1)-E(1,1).
\end{equation}

When the decoy state technique is applied, Eq.~\ref{eq_p} is expanded similar to Eq.~\ref{eq_q}. Take the result of $P_{00}$ under the input of weak coherent state $\mu$ as an example,
\begin{align}
    P_{00,\mu}(a,b) &= P(x_A<-T,x_B<-T|a,b,\mu) \nonumber \\
    &= \sum_{n=0}^{\infty} P(x_A<-T,x_B<-T|a,b,n) \frac{\mu^n}{n!}e^{-\mu} \\
    &\equiv \sum_{n=0}^{\infty} P_{00,n}(a,b) \hspace{2pt} \frac{\mu^n}{n!}e^{-\mu} \nonumber,
\end{align}
where $P_{00,n}$ is the coincidence probability of Fock state $|n\rangle$ with detector settings of $a$ and $b$. To calculate the CHSH violation of state $|0_A 1_B \rangle + |1_A 0_B \rangle$, the $P_{00,1}$ should be extracted similar to Eq.~\ref{eq_Y} as
\begin{equation}
P_{00,1}^{E} =\mu_1 \mu_2 \cdots \mu_L
\sum_{j=1}^{L} \frac{\mu_j^{-2}(e^{\mu_j}P_{00,\mu_j} - P_{00,\mu_0})}{\prod_{1\leq i\leq L,i\neq j} (\mu_i - \mu_j)}.
\end{equation}
The estimation $P_{00,1}^{E}$ is also the upper (lower) bound for $P_{00,1}$, depends on the $L$ is odd (even). The bounds are given by ~\cite{yuan_simulating_2016,valente2017probing}
\begin{equation}
\begin{cases}
     P_{00,1}^{E} - \Delta_L \leq P_{00,1} \leq P_{00,1}^{E} \hspace{8pt} (L \text{ is odd})\\
     P_{00,1}^{E} \leq P_{00,1} \leq P_{00,1}^{E} + \Delta_L \hspace{8pt} (L \text{ is even})
\end{cases},
\end{equation}
with the bound interval as
\begin{equation}
\label{eq_delta}
\Delta_L = (-1)^{L+1} \mu_1 \mu_2 \cdots \mu_L
\left( \sum_{j=1}^{L} \frac{\mu_j^{-2}(e^{\mu_j} - 1)}{\prod_{1\leq i\leq L,i\neq j} (\mu_i - \mu_j)} - 1 \right).
\end{equation}
With all four coincidence probabilities, we can provide an estimation $E^{E}$, an upper $E^+$ and a lower bound $E^-$ for the quantum correlation $E$
\begin{align}
    E^{E}(a,b)&=\frac{P_{00,1}^{E}+P_{11,1}^{E}-P_{01,1}^{E}-P_{10,1}^{E}}{P_{00,1}^{E}+P_{11,1}^{E}+P_{01,1}^{E}+P_{10,1}^{E}}, \\
    E^{\pm}(a,b)&=\frac{P_{00,1}^{\pm}+P_{11,1}^{\pm}-P_{01,1}^{\mp}-P_{10,1}^{\mp}}{P_{00,1}^{\mp}+P_{11,1}^{\mp}+P_{01,1}^{\mp}+P_{10,1}^{\mp}}.
\end{align}
Similarly, for the CHSH violation, we can calculate the estimation $S^{E}$, the upper $S^+$, and the lower bounds $S^-$. 
\begin{equation}
    \begin{split}
        S^{E}=E^{E}(a_0,b_0)+E^{E}(a_1,b_0)\\
        +E^{E}(a_0,b_1)-E^{E}(a_1,b_1) \\
        S^{\pm}=E^{\pm}(a_0,b_0)+E^{\pm}(a_1,b_0)\\
        +E^{\pm}(a_0,b_1)-E^{\mp}(a_1,b_1).
    \end{split}
\end{equation}
Note that the upper and lower bounds do not flip for the last item because it is supposed to be a negative value.

\begin{figure}[tbp]
\centering
\includegraphics[width=0.5\columnwidth]{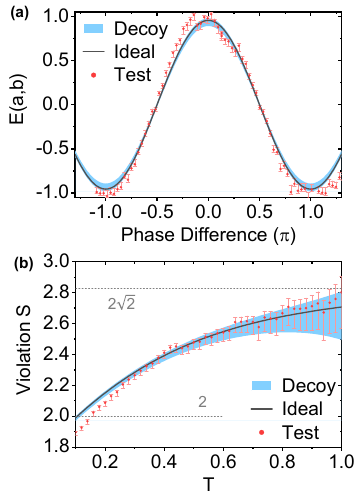}
\caption{(a) The quantum correlation $E$ at different phase difference $d\theta$ between Alice and Bob when $T=1$. (b) The CHSH violation $S$ at different threshold $T$. For both figures, the solid line shows the simulation result considering an ideal single photon input. The blue shade shows the upper and lower bounds simulated using the decoy state technique. The dots are the estimations calculated using experimental data, and the error bars are the upper and lower bounds.}
\label{fig_CHSH}
\end{figure}

For the experiment, the decoy states used are the same as in the previous section. Since the quantum correlation $E(a,b)$ does not depend on the global phase because of global phase drift, the correlation is first evaluated as a function of the phase difference $d\theta$ between Alice and Bob. As shown in Fig.~\ref{fig_CHSH}a, considering a threshold $T=1$, the solid line shows the simulated result with real single photon input using Eq.~\ref{eq_E}, while the blue shade marked the upper and lower bounds simulated using decoy-state method. The dots and the error bars show the experimental estimated value and bounds, respectively. The experimental results exhibit quantum interference, with amplitude bound by [0.973, 1.071]. 

The CHSH result is further calculated as shown in Fig.~\ref{fig_CHSH}b. Higher $T$ leads to a higher correlation and lower success probability. When $T>0.15$, CHSH violation $S>2$, and the Bell inequality is violated. The CHSH lower bound reaches a maximum of $S^{-}=2.59$ at $T = 0.82$.

\section{\label{sec:con} Conclusions} 
In this manuscript, we propose and demonstrate an integrated photonic platform with quantum entanglement generation and witnessing. The single-photon entangled states are generated at a sample speed of 8.75 GS/s. The entanglement is verified first by performing a state tomography that shows a fidelity of 92\%. Then, a CHSH violation test with a lower bond of 2.59 further witnesses the quantum entanglement. In addition, we provide proof of the fair sampling condition to ensure that our experimental results accurately reflect the presence of quantum entanglement in our integrated system. 

Comparing to other common photonic platforms, such as ultra-low-loss silicon nitride and high electro-optic coefficient lithium niobate, our results highlight that silicon photonics, despite its higher loss, offers unique advantages in CMOS-compatible, high-speed homodyne detection and scalable on-chip integration. A missing component for full chip integration is the on-chip laser, but it has already been demonstrated in major photonic foundries using monolithic, heterogeneous, and hybrid integration methods~\cite{tan2023foundry}. 

By introducing an innovative loss-equivalent analysis method, we treat detector inefficiency and electronic noise as equivalent optical attenuation in the source. This optical attenuation is then compensated by increasing the input optical power. Consequently, the stringent requirements on detector design, such as high efficiency and low noise, are significantly reduced. This leads to near-ideal quantum entanglement witnessing on a chip that can operate at high speeds. However, also due to the relatively high loss on chip and the absence of a quantum channel, the method cannot be directly applied to quantum key distribution applications. The loss of chip component are provided in Supplementary Information~S4.

We remark that our focus in this paper is not a loophole-free test of single-photon entangled state but rather on-chip quantum entanglement generation and witnessing based on practical devices and reasonable assumption.

One of the key assumptions in our system is that the source is a phase-randomized coherent state with precise intensity control. For the current proof-of-principle purpose, the phases in our experiment are not randomized perfectly using a phase modulator and a true random number generator. However, it can be achieved easily by operating a semiconductor laser diode in the gain-switching mode~\cite{yuan_robust_2014,kobayashi_evaluation_2014} and applying proper intensity modulations. Additionally, we have considered only the asymptotic case with accurate sources. The statistical fluctuation and the fluctuation of the input state are not included in the analysis but are achievable in recent works~\cite{wang2007decoy, zhu2017parameter, lu2021intensity}. Another assumption required is that the input states are i.i.d. However, our method to boost the sample rate by using a higher detection speed unavoidably introduces some correlation inside each modulation level. One way to solve this is using the same modulation and detection speed. Due to the equipment limitation, our result still shows some intersymbol correlation. 
There are also methods to remove the i.i.d requirement \cite{nahar2023imperfect}, which we leave as future work.       

The quantum entanglement generated in our system is probabilistic and unable to operate in a single-shot manner. Although single-particle entangled states are shown to be useful in quantum communication~\cite{azzini2020single} and quantum sensing~\cite{brown2023interferometric,kolavr2007path,higgins2007entanglement} applications, most applications require deterministic state generation, which is beyond the probabilistic scheme demonstrated here. Therefore, it can be useful in systems where statistical or entropy analysis is sufficient, such as the fundamental study of quantum information processing systems and quantum random number generation \cite{piveteau2024optimization}. These results establish a viable path toward fully integrated, high-bandwidth, room-temperature quantum photonic systems.

\medskip

\begin{backmatter}

\bmsection{Funding}
The authors acknowledge funding support from the Zhejiang Provincial Natural Science Foundation of China (Z25F050014), National Natural Science Foundation of China (NSFC) (62505275), Zhejiang University Education Foundation Qizhen Scholar Foundation, National University of Singapore Start-Up Grant (FY2023), the National Research Foundation of Singapore (NRF) Fellowship grant (NRFF11-2019-0001), NRF Quantum Engineering Programme 1.0 grant (QEP-P2), and NUS Microelectronics Trailblazing Funding (A-8002620-01-00).

\medskip

\bmsection{Disclosures}
The authors declare no conflicts of interest.

\medskip

\bmsection{Data availability} 
All of the data that support the findings of this study are available in the main text. Source data are available from the corresponding author on request.

\end{backmatter}

\bibliography{bib}

\end{document}


\maketitle

\section{Proof of Fair Sampling Condition} 

In our system, the statistics used for the Bell-CHSH test are post-selected based on the threshold $T$ in the quadrature measurement. This raises an essential question: Does the post-selected Bell violation serve as a faithful representation of the quantum system in question and an accurate indicator for the presence of quantum entanglement? 

In this section, we follow the work by Orsucci et al. \cite{orsucci2020post} and verify that the postselection performed in our system does not violate the fair sampling condition. It is important to note that our motivation fundamentally diverges from that of loophole-free Bell inequality violation. Our objective here is to witness quantum entanglement on a chip based on our modeling of the trusted device. 

Based on Ref.~\cite{orsucci2020post}, the post-selected data from a practical measurement device $\mathcal{M}$ can be described as a filter $\mathcal{F}$ that acts on the input quantum state $\rho \in S(\mathcal{H})$ and the classical settings $a \in A$, followed by an ideal lossless detector $\overline{\mathcal{M}}$. 
$S(\mathcal{H})$ is the class of positive semi-definite unit-trace operators on a Hilbert space $\mathcal{H}$. 
A flag $f \in \{\checkmark,\varnothing \} $ is returned by the filter, with the probability depending on both the inputs $a$ and $\rho$. Only when $f = \checkmark$, the ideal detector keeps its output. 

This measurement scheme satisfies the (weak) fair-sampling assumption if there exists a decomposition $\mathcal{M} = \overline{\mathcal{M}} \circ \mathcal{F}$, where the filter $\mathcal{F}$ factories in two parts. The classical part $\mathcal{F}_C : S(\mathcal{H}_A) \rightarrow S(\mathcal{H}_f \otimes \mathcal{H}_A)$ acts on the classical setting $a$, where $\mathcal{H}_A$ is the Hilbert space with an orthonormal basis $\{ \ket{a} \}_{a \in A}$ and $\mathcal{H}_f = \textrm{span}(\ket{\varnothing} , \ket{\checkmark})$. The quantum part $\mathcal{F}_Q : S(\mathcal{H}) \rightarrow S(\mathcal{H}_f \otimes \mathcal{H})$ acts on the quantum input $\rho$. In other words, we require:
\begin{equation}
\label{eq_proof}
    \mathcal{F}(\ket{a} \!\! \bra{a} \otimes \rho) = \wedge [ \mathcal{F}_C(\ket{a}\!\! \bra{a}) \otimes \mathcal{F}_Q(\rho) ],
\end{equation}
where the function $\wedge$ (logical AND) acts only on the flags, and it means that the filter $\mathcal{F}$ returns $\checkmark$ if and only if both $\mathcal{F}_C$ and $\mathcal{F}_Q$ return $\checkmark$. This amounts to the case where the measurement device acts independently on the classical and quantum inputs it receives, conditioned on having had a successful detection~\cite{orsucci2020post}. 

To prove fair sampling assumption hold in our scheme, the POVM of our measurement device is first described as $\{ M_x:=\sum_{a \in A}  \ket{a}\!\! \bra{a} \otimes M_x^a\}$ , where $x$ is a set of possible outcomes. Considering the threshold $T$ applied on the detector outputs, the operator can be separated as 
\begin{equation}
\begin{split}
\label{eq_appen_M}
    M_\varnothing =& \ket{a}\!\! \bra{a} \otimes \int_{-T}^{T} Q^a(x) \,\,dx  \\
     =& \ket{a}\!\! \bra{a} \otimes \widehat{Q}_\varnothing \\
    M_\checkmark =& \ket{a}\!\! \bra{a} \otimes (\int_{-\infty}^{-T} Q^a(x) \,\,dx + \int_{T}^{\infty} Q^a(x) \,\,dx ) \\
    =& \ket{a}\!\! \bra{a} \otimes \widehat{Q}_\checkmark, \\
\end{split}
\end{equation}
where $Q^a(x)$ is the wavefunction of the eigenstate corresponding to measurement setting $a$ and quadrature value $x$. Here, we impose that the measurement only acts non-trivially in the $\{\ket{0}, \ket{1}\}$ Fock states subspace, which is reasonable because the multi-photon components are removed by the decoy-state method. Hence, the operator $\widehat{Q}_\varnothing$ can then be written as
\begin{equation}
\begin{split}
\label{eq_appen_Q}
    \widehat{Q}_\varnothing = \quad & \int_{-T}^{T} \braket{0|x}\!\braket{x|0} \,dx \ket{0}\!\! \bra{0} \\
    +& \int_{-T}^{T} \braket{0|x}\!\braket{x|1} \,dx \ket{0}\!\! \bra{1} \\
    +& \int_{-T}^{T} \braket{1|x}\!\braket{x|0} \,dx \ket{1}\!\! \bra{0} \\
    +& \int_{-T}^{T} \braket{1|x}\!\braket{x|1} \,dx \ket{1}\!\! \bra{1}, 
\end{split}
\end{equation}
with 
\begin{equation}
\begin{split}
    \braket{x|0} =& (\frac{1}{\pi})^{\frac{1}{4}} e^{-1/2 x^2}, \\
    \braket{x|1} =& (\frac{1}{\pi})^{\frac{1}{4}} e^{-1/2 x^2+ i\theta} \sqrt{2} x, 
\end{split}
\end{equation}
where $\theta$ is the local oscillator phase and corresponds to
the measurement setting $a$. Since the $\braket{x|1}$ term is an odd function, the 2nd and 3rd term of Eq. \ref{eq_appen_Q} becomes 0. The operator $\widehat{Q}_\varnothing$ does not depend on the local oscillator phase $\theta$ as well as the measurement setting $a$. The same property also applies for $\widehat{Q}_\checkmark$.  

The filter is defined with shorthand $\xi := \ket{a}\!\! \bra{a} \otimes \rho$ as
\begin{equation}
\begin{split}
\label{eq_appen_filter}
    \mathcal{F}(\xi) = \quad & \ket{\checkmark}\!\! \bra{\checkmark} \otimes \sqrt{M_\checkmark} \; \xi \, \sqrt{M_\checkmark} \\
    +& \ket{\varnothing}\!\! \bra{\varnothing} \otimes \sqrt{M_\varnothing} \; \xi \, \sqrt{M_\varnothing}. \\
    = \quad & \ket{\checkmark}\!\! \bra{\checkmark} \otimes \ket{a}\!\! \bra{a} \otimes \sqrt{\widehat{Q}_\checkmark}  \; \rho \, \sqrt{\widehat{Q}_\checkmark} \\
    +& \ket{\varnothing}\!\! \bra{\varnothing} \otimes \ket{a}\!\! \bra{a} \otimes \sqrt{\widehat{Q}_\varnothing} \; \rho \, \sqrt{\widehat{Q}_\varnothing}.
\end{split}
\end{equation}

Now, note that in our system, data are only discarded based on the quadrature value $x$ and not the measurement setting $a$. Since we have shown that the operator $\widehat{Q}$ does not depend on the measurement setting $a$, we can write down the following decompositions:

\begin{align}
    \mathcal{F}_C(\ket{a}\!\! \bra{a}) = \; &\ket{\checkmark}\!\! \bra{\checkmark} \otimes \ket{a}\!\! \bra{a} \\
    \mathcal{F}_Q(\rho) = \; & \ket{\checkmark}\!\! \bra{\checkmark} \otimes \sqrt{\widehat{Q}_\checkmark} \rho \sqrt{\widehat{Q}_\checkmark} \notag\\
    +& \ket{\varnothing}\!\! \bra{\varnothing} \otimes \sqrt{\widehat{Q}_\varnothing} \rho \sqrt{\widehat{Q}_\varnothing},
\end{align}
which indeed describes the postselection process via thresholding of the quadrature measurement in our device. According to our modeling, Eq.~\ref{eq_proof} clearly holds and confirms that the (weak) fair sampling assumption remains intact, although the postselection via thresholding of quadrature measurement was performed if the input quantum state is a linear operator in the $\{\ket{0}, \ket{1}\}$ Fock states subspace. 

Hence, this shows that post-selected Bell violation, when used appropriately, can be a precious tool for unveiling quantum entanglement that is ``hidden'' in the mixture with separable states. In our case, by employing the decoy-state method, we could bound the proportion of quantum state measured that are residing in the $\{\ket{0}, \ket{1}\}$ Fock states subspace as well as its post-selected Bell violation, which indicates the amount of entanglement present. Such a technique may prove valuable in the applications of quantum communications. 

\section{System Stability Analysis}
\begin{figure}[tbp]
\centering
\includegraphics[width=0.7\columnwidth]{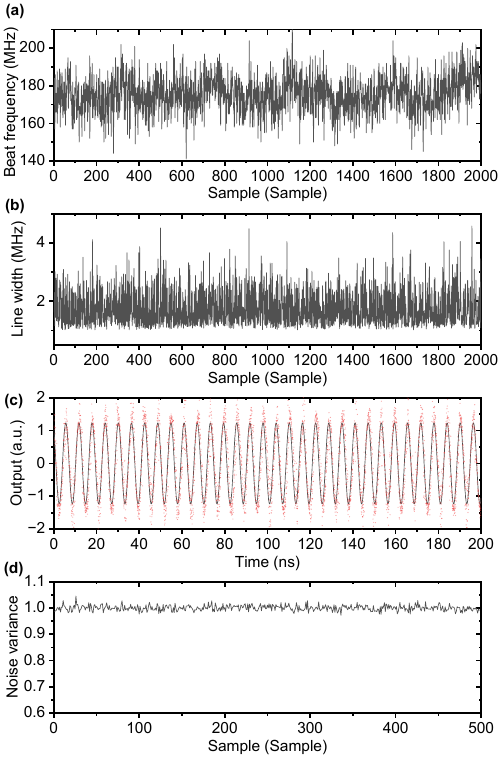}
\caption{(a) Beat frequency between signal and LO. (b) Laser linewidth extracted by beating output. (c) Beat output in time domain (dots) with the ideal fitted beat output (line). (d) Measured shot noise variance throughout the experiment. }
\label{fig_phase}
\end{figure}

To avoid the low-frequency cut-off of our system and satisfy the phase randomization requirement of decoy state techniques, the LO and signal are generated from different lasers with about 170 MHz detuning. The phase drifts on different time scales are shown in Fig.~\ref{fig_phase} (a-c). In the minute scale, Fig.~\ref{fig_phase} (a-b) shows the laser detuning and laser linewidth drift, recorded approximately every 2 s per sample. The beat frequency drifted randomly by about 50 MHz on the minute time scale. The laser linewidth is measured to be about 1 to 3 MHz. Fig.~\ref{fig_phase} (c) shows the time domain beat signal on the nanosecond scale compared to the ideal sine fit. The deviation clearly shows a random phase drift, especially in the 70-100 ns window, with $0.05 \pi$ clearly shown, confirming the presence of rapid, stochastic phase variations. The results confirmed that randomization is effective (but not i.i.d.) for our experiment.

Moreover, to confirm the system stability, the measured shot-noise variance over an extended experimental period (approximately one day) is shown in Fig.~\ref{fig_phase} (d). The exact time stamps were not recorded during the experiment. Each sample in the figure is collected roughly every 1 min. The results indicate that the variance remains stable throughout the measurement sequence.

\section{Electromagnetic Interference Issue}
\begin{figure}[tbp]
\centering
\includegraphics[width=0.7\columnwidth]{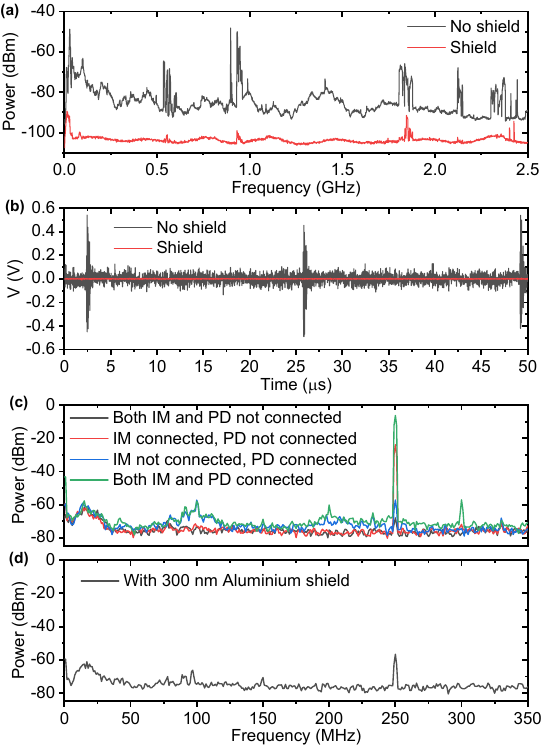}
\caption{(a-b) The observed environmental electromagnetic interference signal with and without proper shielding is shown in (a) frequency domain and (b) time domain. (c) Chip crosstalk between the intensity modulator (IM) and photodiodes (PD). The testing verifies the contribution from IM and PD. (d) The effect on the crosstalk with 300 nm aluminum shield applied on the chip surface. }
\label{fig_emi}
\end{figure}

A main issue in the experiment is the electromagnetic interference (EMI) from both the environment source and chip crosstalk. Due to the high gain amplifier (39~dB) used in the homodyne detector, the tiny amount of EMI captured by the photodiode electrodes is not negligible in the amplifier output spectrum. Fig.~\ref{fig_emi}a shows the environmental EMI noise in the 0-2.5~GHz band, which is the most noisy band that contains the noise from radios, mobile phones, WiFi, Bluetooth devices, etc. The time domain signal is shown in Fig.~\ref{fig_emi}b, where some unknown spikes can be observed. To reduce the environmental EMI, the amplifiers are sealed in an aluminum box with minimum openings only for the input and output cables. All the cables are changed to shielded cables. The uncovered parts are carefully wrapped with conductive tapes. With proper shielding in place, the resulting frequency and time signals are shown as red curves in Fig.~\ref{fig_emi}a-b. A more than 40 dB noise suppression is observed.   

Chip crosstalk between the modulators and homodyne detectors is another EMI source. The effect is observed by applying a 250~MHz 2~Vpp signal on the IM with the light off and observing the response on the PD. Fig.~\ref{fig_emi}c shows four curves when probes connect both the IM and PD, each of them is connected, and both are disconnected. The result proves that the probes introduce minimum crosstalk down to -68~dBm. The main contributions are the electrodes on the IM and PD, which show a peak noise of up to -6.3~dBm. To solve this, a 300 nm thick aluminum layer is deposited on top of the chip to make all the electrodes a stripline-like structure. All the unused electrodes on the chip are grounded to use as guard traces. The result is shown in Fig.~\ref{fig_emi}d, demonstrating a crosstalk suppression of 50.3~dB. To further reduce the crosstalk, a smaller driving voltage of 0.5~Vpp is chosen to reduce the effect further.

\section{Chip Component Characterisation} 
\begin{figure}[tbp]
\centering
\includegraphics[width=0.8\columnwidth]{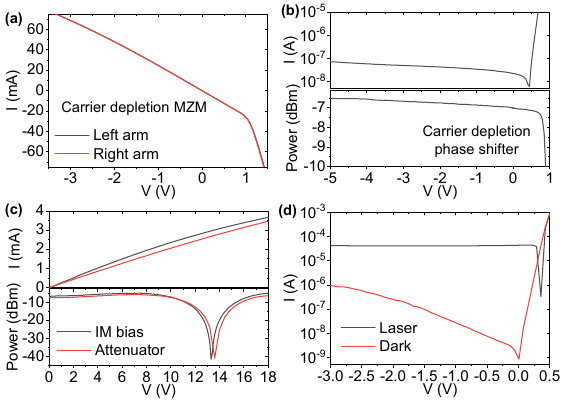}
\caption{(a) I-V curve for the carrier depletion MZM. (b) I-V curve and optical power-V curve for the carrier depletion phase modulator. (c) I-V and optical power-V curves for the thermo-optical phase shifters in the intensity modulators for bias control and the variable optical attenuators. (d) I-V curve for photodiodes on-chip with and without laser input. }
\label{fig_chip}
\end{figure}

More details for the chip components are provided and shown in Fig.~\ref{fig_chip}. The MZMs are 50 $\Omega$ terminated, which can be observed from the I-V curve shown in Fig.~\ref{fig_chip}a. The result also tells us the biasing direction of the P-N junction and the linear region we can use for modulation. For the carrier injection or depletion modulators, the phase modulation is related to the loss induced, which can be measured from a phase modulator with the same structure as MZM. As shown in Fig.~\ref{fig_chip}b, the I-V curve and optical power-voltage dependency show that for a 1.5~mm length phase modulator, the estimated phase-dependent loss is about 0.35~dB for a $\pi$ shift.

\begin{table}[tbp]
\def\arraystretch{1.4}
    \caption{Loss or efficiency of chip component}
    \centering
    \begin{tabular}{l l l}
    \hline\hline 
         Component  &  Loss /   efficiency  & Data   source \\
    \hline
        Waveguide                & 1.40   dB/cm         & Handbook       \\
        Beam   splitter (MMI)    & 0.08 dB              & Handbook       \\
        Photodiode               & \textgreater{}61.7\% & Measured       \\
        Electronic   noise $\eta_{ele}$ & \textgreater{}60\%   & Measured \\  
    \hline\hline     
    \end{tabular}
    \label{tab_loss}
\end{table}

Thermo-optical phase shifters are another component used for low-speed phase tuning, especially in the MZI-based variable optical attenuators and the MZM modulators for modulation point selection. The same phase shifters in the attenuators and MZMs are measured by monitoring the output optical power after the interferometer. The I-V and power-V curves are shown in Fig.~\ref{fig_chip}c. With some BS imbalance, the result shows a $V_\pi$ of about 13.5~V.  

The I-V of photodiodes are also tested as shown in Fig.~\ref{fig_chip}d. The dark current and photocurrent under about 10~dBm laser input are shown. The reverse bias voltage of 1~V is chosen in our experiment, which gives us $2.8\times10^{-8}$~A dark current. 

Although the on-chip loss is not critical in our protocol, we provide the loss values as a reference. Table~\ref{tab_loss} shows the estimated losses for all relevant chip components. Some of the data are directly collected from the foundry handbook.

\section{Quantum State Tomography} 
\begin{figure}[tbp]
\centering
\includegraphics[width=0.9\columnwidth]{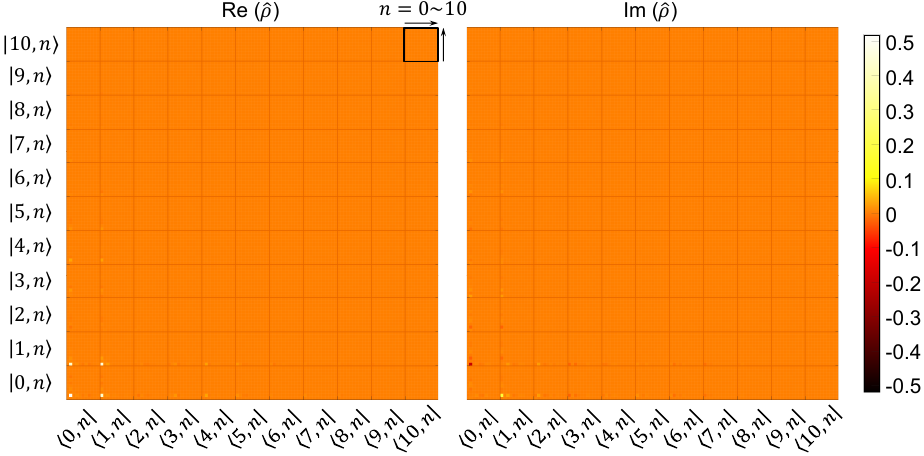}
\caption{The density matrix’s real and imaginary parts with up to 10 photons.}
\label{fig_tomo}
\end{figure}

The tomography was performed using a maximum-likelihood reconstruction with a 10-photon cutoff. Given the mean photon number of 0.984, the probability of more than 10 photons is $8.5 \times 10^{-9}$, which is negligible. For completeness, we have now included the higher-order photon-number terms in Fig.~\ref{fig_tomo}. All contributions beyond two photons are below $2.9 \times 10^{-2}$, confirming their negligible impact and fidelity of our generated entanglement. The remaining non-idealities are primarily due to imperfect modulation and detection (e.g., non-flat spectrum response). These imperfections can be mitigated in future work using advanced digital signal processing techniques such as equalization, filtering, and pulse shaping.

\bibliography{sample}
